\documentclass[aps,physrev,twocolumn,amsmath,a4paper,amssymb,longbibliography,10pt]{revtex4-2}
\usepackage{natbib}
\usepackage{graphicx}
\usepackage{CJK}
\usepackage{color}

\newcommand{\cmmtb}{}

\def\Wcm2{W/cm$^2$}

\begin{document}
\begin{CJK*}{UTF8}{gbsn}

\title{\cmmtb Strong-Field-Assisted X-ray Second Harmonic Generation}

\author{Xinhua Xie (谢新华)}
\email[Electronic address: ]{xinhua.xie@psi.ch}
\affiliation{SwissFEL, Paul Scherrer Institute, 5232 Villigen PSI, Switzerland}

\date{\today}

\begin{abstract}
{\cmmtb We theoretically demonstrate x-ray second harmonic generation (SHG) with strong-field-driven high harmonic generation (HHG) from core electrons in gas-phase atoms}, highlighting its potential as a powerful spectroscopic tool for studying atomic and molecular systems. Our simulations confirm the generation of HHG with x-ray SHG (HHG-XSHG) from atomic core electrons through the rescattering of electrons from x-ray two-photon excitation and subsequent tunneling ionization in an optical laser field. The resulting HHG-XSHG spectrum features a broadband multi-peak structure and a distinct spectral cutoff in the x-ray regime. These findings indicate that HHG-XSHG is a valuable technique for probing core-electron dynamics, generating attosecond x-ray pulses, and exploring nonlinear interactions, effectively merging laser-driven attosecond technology with nonlinear x-ray methodologies provided by x-ray free-electron lasers.
\end{abstract}

\maketitle
\end{CJK*}

\section{Introduction}

Attosecond technology based on laser-driven high harmonic generation (HHG) has emerged as a powerful tool due to its attosecond temporal resolution \cite{Corkum2007}. Over the past decades, it has demonstrated numerous applications in gas-phase atomic, molecular physics, and condensed matter studies \cite{Li2020,Heide2024}. However, one limitation of laser-driven HHG sources is the restricted photon energy due to low conversion efficiency in the x-ray regime. This limitation has urged the development of methods to extend the capabilities of HHG in the x-ray regime, where higher photon energies can offer deeper insights into core-electron dynamics and other fundamental processes.

Simultaneously, the advent of x-ray free-electron lasers (XFELs) has revolutionized nonlinear spectroscopy, including second harmonic generation (SHG), by extending these techniques into the x-ray regime. XFELs offer unprecedented opportunities for material characterization with atomic precision, enabling the probing of core-level electronic transitions and providing element-specific information
with atomic-scale spatial resolution \cite{Prat2020,Habib2023,Mcneil2010,Prat2023,Sobolev2020,Mcneil2010,Shwartz2014,Lam2018}. This breakthrough has paved the way for studying complex systems, including nanostructures, multilayered materials, and materials with strong electron correlations and magnetic properties. The intense pulses from XFELs further facilitate the investigation of nonlinear
interactions and higher-order effects, positioning x-ray SHG as a transformative approach for advancing research in material science, condensed matter physics, and nanotechnology.

Since the invention of the laser, SHG spectroscopy and microscopy have evolved into highly sensitive nonlinear optical techniques widely used for material studies \cite{Boyd2003}. By converting incident light into a new frequency that is double the original, SHG offers a powerful tool of probing surfaces, interfaces, and non-centrosymmetric systems inherently lacking inversion symmetry.
SHG has found broad application across various fields, including surface science, biological imaging, and the characterization of nonlinear materials \cite{Eisenthal1996,Heinz1982,Wang2019,Corn1994,Campagnola2002}. {\cmmtb Previous studies have laid a strong foundation for understanding SHG in various systems. Early experimental work by Fiebig et al. \cite{Fiebig2005} and Kirilyuk et al. \cite{Kirilyuk2005} explored SHG in magnetically ordered and centrosymmetric materials, demonstrating how nonlinear optical effects can be leveraged to probe material properties. On the theoretical side, Pavlyukh et al. \cite{Pavlyukh2012}, Draxl et al. \cite{Sharma2004}, and Andersen et al. \cite{Andersen2002} developed models for nonlinear susceptibilities, electron correlations, and selection rules governing SHG processes. These pioneering efforts have significantly influenced modern developments in x-ray nonlinear optics and continue to serve as a foundation for advancing attosecond spectroscopy in the x-ray regime.} Moreover, time-resolved SHG techniques with ultrafast temporal resolution provide vital insights into rapid dynamical processes, making them indispensable for studying ultrafast structural changes and dynamics \cite{Schroder1991,Tyson2018,Hohlfeld1995,Chang1997,Shang2001,Sinha2023}. {\cmmtb The advent of XFELs has enabled the first experimental demonstrations of x-ray SHG in crystalline solids, where inversion symmetry is naturally broken \cite{Shwartz2014,Lam2018}. These studies have provided insights into electronic correlations and interfacial properties with high spatial and temporal resolution \cite{Glover2012,Sobolev2020}. Recent advancements have further extended x-ray SHG to complex material systems, such as multilayer structures and strongly correlated materials \cite{Shwartz2014,Zhu2021}.} However, achieving x-ray SHG in randomly oriented gas-phase atoms and molecules poses significant challenges due to the lack of global inversion symmetry. As a second-order nonlinear process, SHG is forbidden in centrosymmetric environments \cite{Boyd2003}, and to date, there have been no experimental reports of x-ray SHG in such systems.

\begin{figure}[ht]
\centering
\includegraphics[width=0.5\textwidth,angle=0]{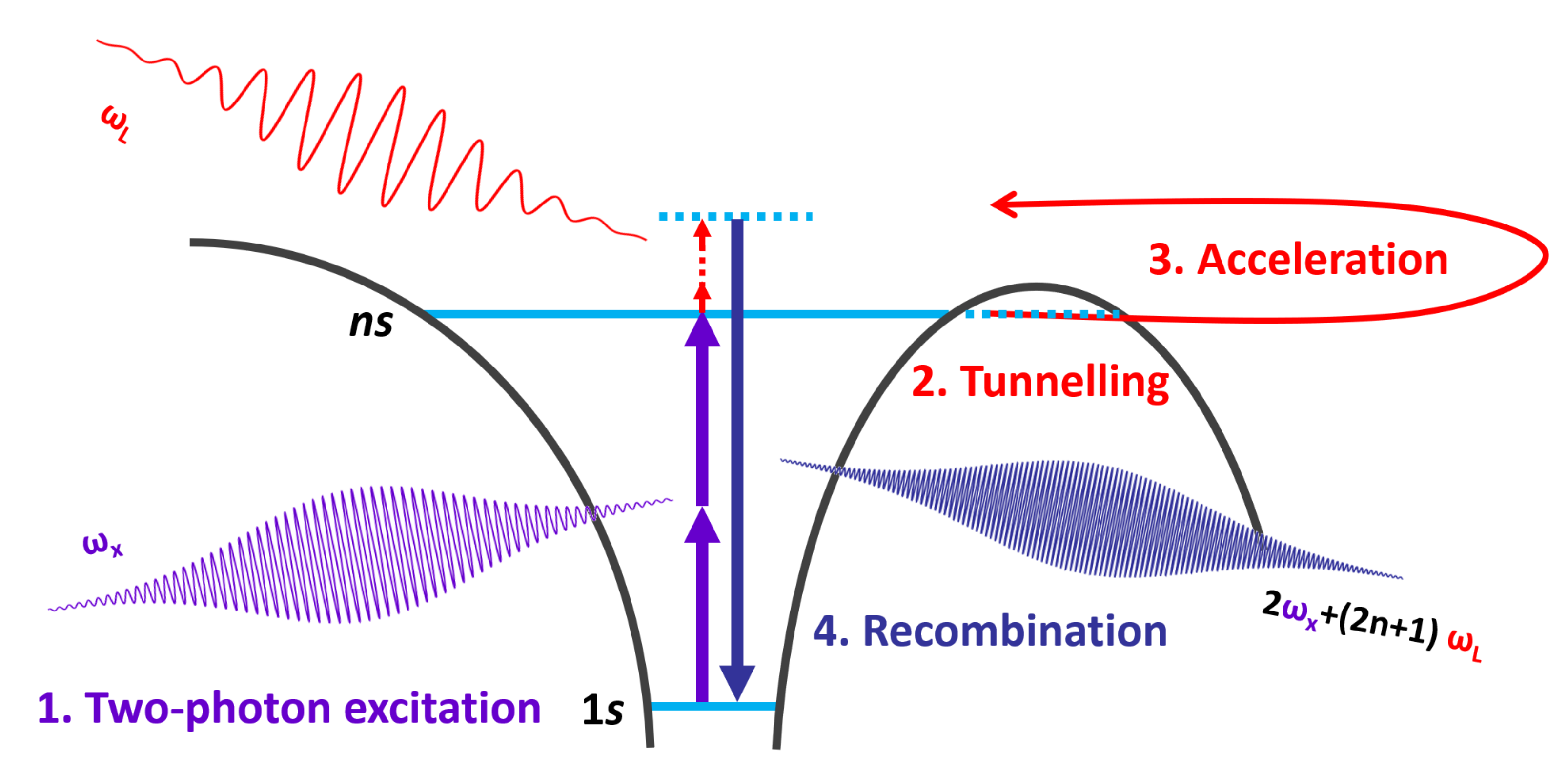}
\caption{Schematic of the HHG-XSHG process. The 1s core electron of an atom is excited to an unoccupied ns state through the absorption of two x-ray photons. The electron in the ns state is driven by a near-infrared laser field, which enables recombination back to the $1s$ state, resulting in the emission of a photon with energy approximately twice that of the incoming x-ray photons.} \label{fig:peda}
\end{figure}

To address these limitations in HHG and x-ray SHG, we propose a method that combines laser-driven HHG with x-ray SHG (HHG-XSHG). This HHG-XSHG method uses an intense optical laser field along with an x-ray pulse to interact with core electrons in an atom, as illustrated in Fig.~\ref{fig:peda}. This combined external field enables nonlinear interactions, facilitating the generation of HHG-XSHG. The optical laser perturbs the electron cloud populated through x-ray two-photon excitation, creating transient dipoles that allow the recombination of the electron with the core hole, thereby enabling HHG-XSHG. This approach extends the potential of HHG and nonlinear x-ray spectroscopy, enabling x-ray SHG in gas-phase systems where it was previously unfeasible. {\cmmtb Additionally, it provides access to dark states that cannot be reached through single x-ray photon processes.}

In this work, we simulate the HHG-XSHG process using the single active core electron approximation by solving the time-dependent Schr\"odinger equation. To enhance the HHG-XSHG signal, we leverage a two-photon excitation resonance, where two x-ray photons excite the system from a core-level state to a valence state. By tuning the x-ray photon energy to match a core-to-valence transition, we significantly enhance the nonlinear interaction, leading to a substantial increase the generation efficiency of HHG in the x-ray regime. This resonant enhancement not only boosts the HHG-XSHG signal but also provides valuable insights into core-electron dynamics and ultrafast processes in gas-phase atoms and molecules.

\section{Method}

The interaction in the HHG-XSHG process primarily involves a single core electron, while the optical laser interacts with the loosely bound electron in the excited states that are promoted during the two-photon excitation process. Due to this focus on a single core electron, we employ the single-active-electron (SAE) approximation for modeling the process. To simulate the HHG-XSHG process of gas-phase atoms, we performed numerical calculations by solving the two-dimensional time-dependent Schr\"odinger equation (TDSE) under the SAE approximation \cite{Broin2014,Zhang2023,Bauer2006,Schneider2023,xie07jmo,xie07pra,Xie2015,Deng2015,Zhang2020}.

For these simulations, we employ a neon-like atomic model using a soft-core potential, defined as:
$V(x,y) = -e^{-a\sqrt{x^2+y^2}}/\sqrt{x^2+y^2+b^2}$, where \(a=0.005827\) and \(b=0.305\). The ground 1s state has a binding energy of 870.2 eV, which matches the ionization potential of the neon 1s core electron, while the first excited 2s state has a binding energy of approximately 4.9 eV, closely matching the binding energy of the neon's first excited \(3s\) state.
{\cmmtb The primary decay process of the neon 1s core-hole state is Auger decay, where the vacancy is filled by a higher-energy electron, leading to the emission of a secondary electron. This process occurs on a timescale of approximately 2.4 fs \cite{Coreno1999,Haynes2021}.
To account for the finite lifetime of the neon 1s core-hole state, we introduce a decay term in the TDSE, effectively removing the core-hole state over time. This approach models the Auger decay process by including a complex energy shift, where the imaginary component corresponds to the decay rate \cite{Schlatholter1999,Buth2013}.}
The vector potentials of the linearly polarized x-ray (X) and optical laser (L) pulses are defined as:
$A_F(t) = \frac{\mathcal{E}_F}{\omega_F} A_{F0}(t) \sin{(\omega_F t)}, \quad F = \{\text{X, L}\}$,
where \(\mathcal{E}_F\) and \(\omega_F\) are the peak electric field and center frequency, respectively.
A super-Gaussian envelope is exploited for both the x-ray and optical laser pulses: $ A_{F0}(t) = \exp{\left(-2\ln2\frac{t^{12}}{\tau_F^{12}}\right)}.$
The duration of the optical laser pulse, \(\tau_L\), is set to 5 fs, resulting in a flattop shape with approximately three and a half optical cycles in its plateau. The x-ray pulse shares the same envelope shape and is temporally overlapped with the optical laser pulse.

To analyze the interaction between the model atom and the external fields, we calculate the power spectrum of the dipole radiation by taking the Fourier transform of the dipole acceleration over the propagation time. This allows us to extract the characteristics of the generated second harmonic signal and understand the nonlinear response of the system.

\section{Results and discussion}

\subsection{Simulated HHG-XSHG spectra}

\begin{figure}[ht]
\centering
\includegraphics[width=0.5\textwidth,angle=0]{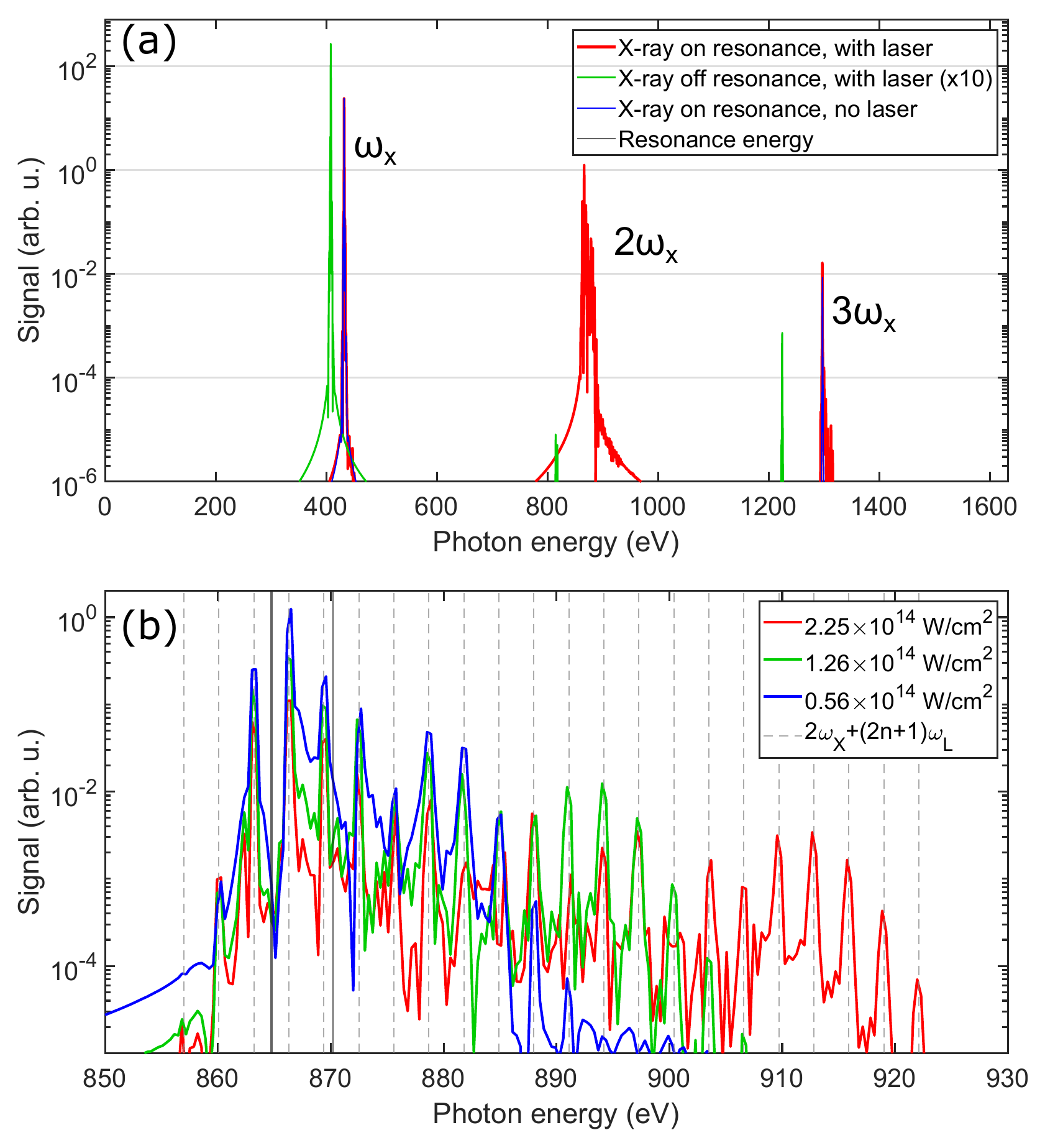}
\caption{(a) Simulated harmonic spectra with different conditions: two-photon resonance with only x-ray (blue), two-photon resonance with both x-ray and optical laser (red), below two-photon resonance with both x-ray and optical laser (green). (b) Simulated x-ray second harmonic spectra for three optical laser intensities with the x-ray at two-photon resonance. The dashed vertical lines indicate the photon energy of $2\omega_X+(2n+1)\omega_L$, and the thick gray solid vertical line is at $2\omega_X$. The thin solid line presents the ionization potential at 870.2 eV. } \label{fig:spec}
\end{figure}

Figure~\ref{fig:spec}(a) presents the simulated dipole radiation spectra obtained from the TDSE calculations. The red curve depicts the HHG-XSHG signal when the x-ray photon energy is tuned to 432.7 eV, corresponding to the two-photon resonance for the 1s to 2s transition (865.4 eV) in the neon-like model. The {\cmmtb green} curve shows the SHG signal with the x-ray photon energy off-resonance at 408 eV, while the {\cmmtb blue} curve represents the spectrum in the absence of the optical laser. In these simulations, the optical laser has a central wavelength of 800 nm (1.55 eV) and a peak intensity of $0.56 \times 10^{14} \ \text{W/cm}^2$, interacting with the system alongside the x-ray pulse at a peak intensity of $3.5 \times 10^{16} \ \text{W/cm}^2$, achievable with modern XFEL facilities \cite{Prat2020,Habib2023,Mcneil2010,Prat2023,Sobolev2020}.

The simulations yield several key observations. The green curve, representing the scenario without the optical laser, shows no second harmonic signal, in agreement with selection rules that prohibit SHG in centrosymmetric systems. When the x-ray photon energy is below the two-photon resonance ({\cmmtb green} curve), a weak second harmonic signal appears, but it remains about three orders of magnitude lower than the third harmonic signal. In contrast, when the x-ray photon energy is tuned to the two-photon resonance (red curve), the second harmonic signal is significantly amplified, exceeding the third harmonic signal by about two orders of magnitude. This enhancement is attributed to the increased probability of two-photon absorption, emphasizing the critical role of resonance in enhancing x-ray SHG efficiency. These results demonstrate that the laser-assisted approach can greatly enhance SHG signals, even in inherently centrosymmetric systems, under optimal resonance conditions.

Figure~\ref{fig:spec}(b) provides a closer examination of the HHG-XSHG spectrum, focusing on the photon energy range between 850 and 930 eV for three different optical laser intensities. The spectra exhibit a broad, multi-peak structure characterized by an extended plateau and a sharp cutoff, reminiscent of features observed in laser-driven high harmonic generation (HHG) \cite{Li1989,Corkum1993,Lewenstein1994}. According to selection rules, only harmonics involving combinations of two x-ray photons and an {\cmmtb odd} number of optical laser photons contribute to the second harmonic signal. These contributions are indicated by the vertical dashed lines in Fig.~\ref{fig:spec}(b) at photon energies of $2\omega_X + (2n+1)\omega_L$, where $n$ is an integer.

As the optical laser intensity increases, the cutoff extends to higher photon energies, indicating that the electron wave packet, initially excited through the two-photon transition from the 1s state, is further driven by the optical laser field before recombining with the 1s core hole. During this process, the electron wave packet may undergo additional excitation or tunnel ionization before recombination.

The cutoff energies can be understood using the relation $I_p + 3.17 U_p$, where $I_p$ is the core electron ionization potential and $U_p$ is the ponderomotive potential of the optical laser field \cite{Corkum1993,Lewenstein1994,Corkum2007}.
The spectral plateau corresponds to the acceleration of the electron wave packet by the optical laser field, followed by recombination with the core hole, leading to the emission of harmonics from the combined x-ray and optical fields.

Taking the rescattering picture from laser-driven HHG, we can describe the generation of radiation above the ionization threshold with a four-step mechanism: (1) two-photon excitation, (2) ionization, (3) acceleration, and (4) recombination, as schematized in Fig.~\ref{fig:peda}. For radiation below the ionization threshold, the process resembles below-threshold HHG, where the optical laser facilitates electron excitation between the x-ray-induced two-photon state and the final recombination \cite{Xiong2016,Xiong2014,Xiong2017}.
In this mechanism, the role of the x-ray pulse is primarily to initiate the two-photon excitation, as its photon energy exceeds the binding energy of the valence orbitals. {\cmmtb The optical laser, with its lower photon energy, primarily governs the subsequent excitation or ionization, acceleration, and recombination processes by interacting predominantly with valence electrons and those in the continuum. Notably, below-ionization-threshold harmonics exhibit stronger signals compared to those above the threshold. This is because the recombination cross section between bound states is significantly larger than that between bound and continuum states.} Additionally, the radiation intensity below the second harmonic photon energy (corresponding to $n < 0$) declines rapidly as the photon energy decreases, showing only one prominent peak. This behavior results from the limited probability of the optical laser to drive the electron towards lower binding energies.

It is worth noting that the HHG-XSHG spectrum is broad, extending over 50 eV at an optical laser intensity of $2.25 \times 10^{14} \ \text{W/cm}^2$, despite the fundamental x-ray pulse having a bandwidth of about 0.4 eV. This broad bandwidth implies that if the dispersion of the SHG pulse can be minimized, it could support pulse trains with an individual pulse duration below 40 attoseconds, offering potential for ultrafast time-resolved studies.

\subsection{Time-frequency analysis of HHG-XSHG}

\begin{figure}[ht]
\centering
\includegraphics[width=0.5\textwidth,angle=0]{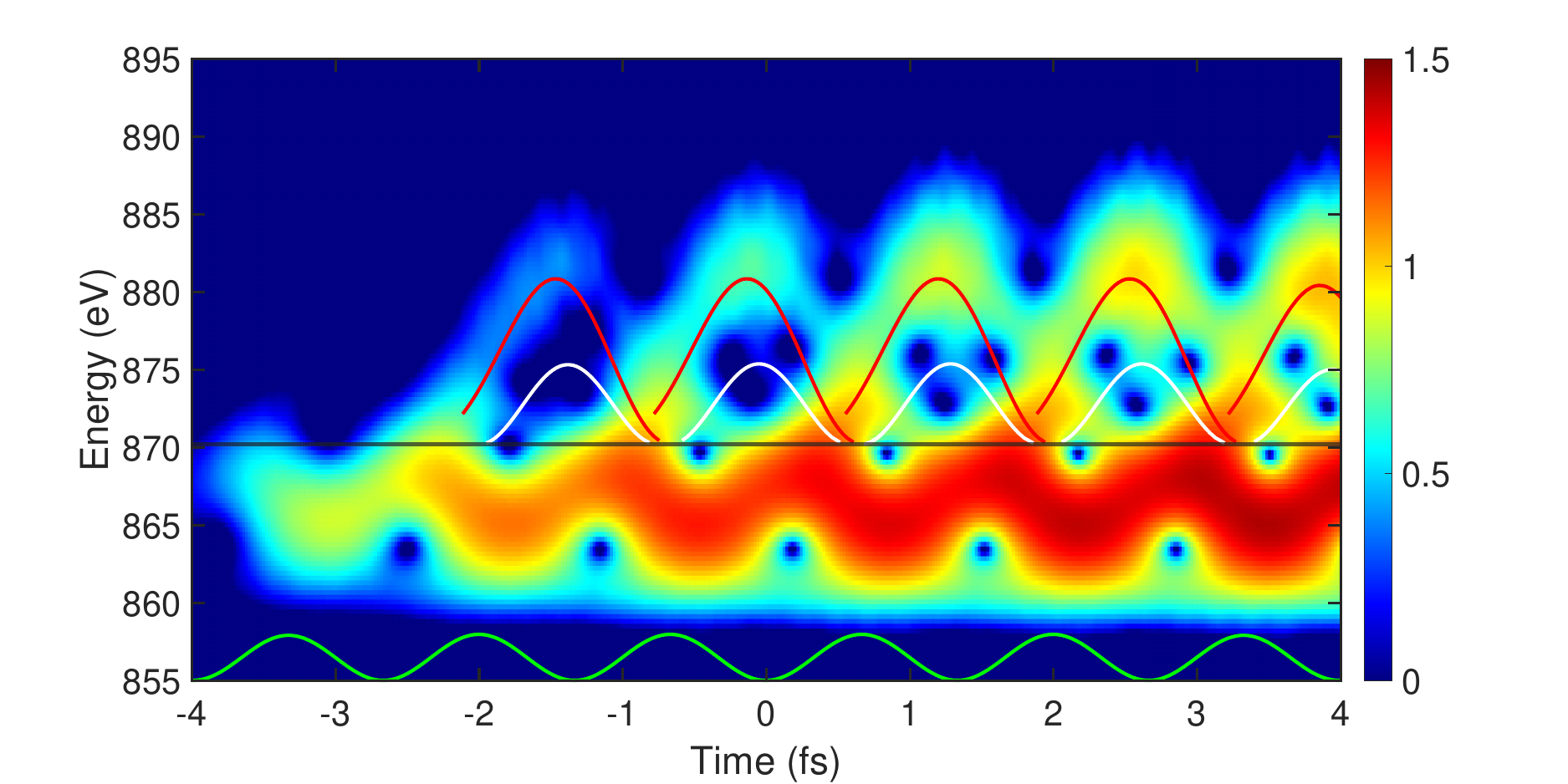}
\caption{Time-frequency analysis of HHG-XSHG with the optical laser intensity of \(0.56 \times 10^{14} \, \text{W/cm}^2\). The red and blue curves represent the relationship between the rescattering energy of the first and second returning electrons and the rescattering time calculated using classical electron trajectories. The green line indicates the intensity of the optical laser field.{\cmmtb The horizontal line at 870.2 eV represents the ionization threshold.}} \label{fig:tfa}
\end{figure}

To gain detailed temporal and spectral insights from the simulated dipole acceleration data, we performed a time-frequency analysis using the windowed Fourier transform. This approach allows us to resolve the time-dependent evolution of the frequency components, providing a deeper understanding of the dynamic processes that govern the dipole evolution. The time-frequency structure of the HHG-XSHG
dipole radiation is presented in Fig.~\ref{fig:tfa}, which reveals the temporal evolution of the signal and elucidates the underlying mechanisms. The time-frequency analysis not only highlights the dynamic nature of the HHG-XSHG process but also identifies specific time intervals where significant radiation contributions arise. This information is critical for optimizing experimental conditions and enhancing SHG efficiency in practical applications.

To interpret the features observed in the time-frequency distributions {\cmmtb above the ionization threshold}, we conducted classical trajectory simulations by solving the Newtonian equations of motion for a free electron in the presence of the optical laser field. These simulations provide the electron's rescattering energy (including the ionization potential) as a function of the rescattering time, shown as red and blue curves in Fig.~\ref{fig:tfa}, corresponding to the first and second electron returns, respectively.
In these simulations, the Coulomb potential was neglected to focus solely on the interaction between the free electron and the laser field. The rescattering energy was calculated at the moment the electron returns to the origin, offering a direct view of the energy dynamics during rescattering events. The red curves, representing the first rescattering events, show a strong correlation with the time-frequency structures {\cmmtb in the ionization threshold regime} that repeat during each half-cycle of the optical field for signals above the ionization threshold, which confirms the rescattering mechanism. These curves also predict the cutoff energy, indicating the maximum energy gained by the electron during rescattering.
Conversely, the periodic structure observed below the ionization threshold corresponds to the dynamics of the laser-driven bound electron. {\cmmtb This region reflects the interaction of the optical field with the bound states populated by the x-ray two-photon excitation, resulting in a modulated features that contribute to the strong below-ionization-threshold HHG-XSHG signal \cite{Xiong2016,Xiong2014,Xiong2017}.}

\subsection{Dependence on laser intensity, x-ray intensity and photon energy}

\begin{figure*}[ht]
\centering
\includegraphics[width=0.9\textwidth,angle=0]{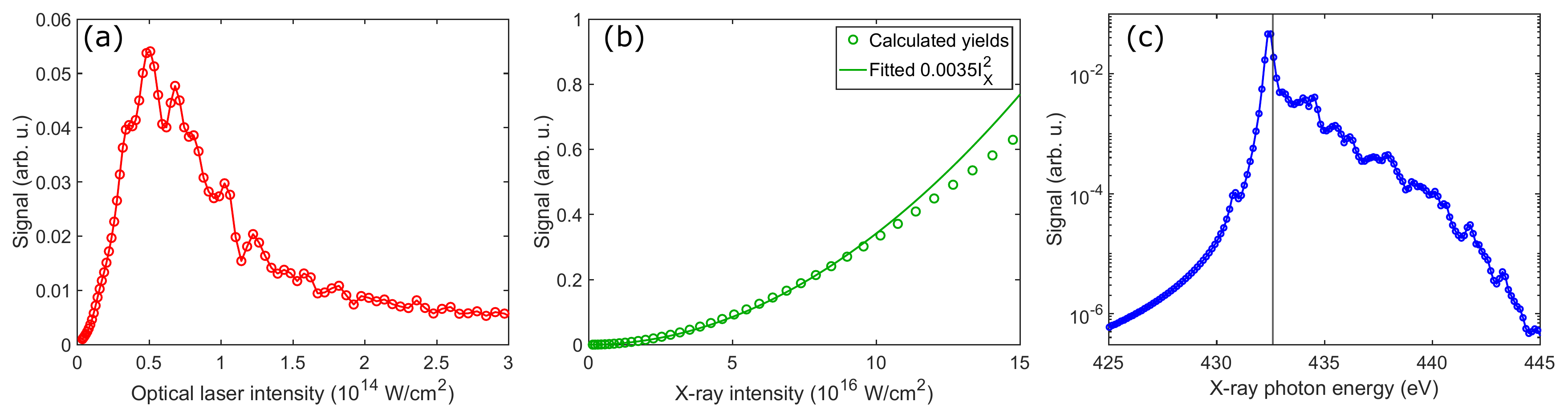}
\caption{ The dependence of the HHG-XSHG on the optical laser intensity (a), the x-ray intensity (b) and the x-ray photon energy (c). {\cmmtb The vertical line in (c) indicates the two-photon resonance energy between the 1s and 2s states.}} \label{fig:depend}
\end{figure*}

To further optimize the HHG-XSHG yield with respect to the x-ray and optical laser parameters, we performed simulations of the HHG-XSHG process, varying the optical laser intensity, x-ray intensity and photon energy.

As shown in Fig.~\ref{fig:spec}(b), the HHG-XSHG spectrum exhibits a significant dependence on the optical laser intensity. We conducted simulations with laser peak intensities up to $3 \times 10^{14} \ \text{W/cm}^2$. The HHG-XSHG yields, integrated over the photon energy range from 850 eV to 1000 eV, are plotted in Fig.~\ref{fig:depend}(a) as a function of the laser peak intensity.
In the low-intensity regime, the SHG yield increases with rising laser intensity, reaching a maximum at around $0.5 \times 10^{14} \ \text{W/cm}^2$. Beyond this point, the HHG-XSHG yield starts to decline and levels off at approximately $2.5 \times 10^{14} \ \text{W/cm}^2$. The initial increase in yield at low intensities can be explained as follows: the x-ray pulse excites the 1s core electron to the 2s state, and the optical laser field then drives the electron from the 2s state into further excitation or tunneling ionization. These processes are highly non-linear, with their probabilities increasing as the laser intensity rises. However, at higher laser intensities, the optical laser can ionize most of the electrons excited by the x-ray pulse.
Additionally, as the laser intensity increases, the laser-driven electron gains higher energy, resulting in the emission of higher-energy photons, as seen in Fig.~\ref{fig:spec}(b). Comparing the spectra at intensities of $0.56 \times 10^{14} \ \text{W/cm}^2$ and $2.25 \times 10^{14} \ \text{W/cm}^2$, we observe that the below-threshold spectrum is stronger at lower intensities, while higher photon energy signals dominate at greater intensities. In essence, as the laser intensity increases, more electrons are promoted to the continuum, leading to stronger high-order signals while the low energy signals weaken.
This decline in HHG-XSHG yield at higher intensities can be attributed to the interplay between electron population in continuum states and the recombination probability, which generally decreases as photon energy increases beyond resonance \cite{Zschornack2007}. The resonance-like peak structures observed in the intensity range from $0.5$ to $1.5 \times 10^{14} \ \text{W/cm}^2$ can be explained by the channel-closing effect in laser-driven excitation \cite{Chetty2020}.

The HHG-XSHG yield is expected to be proportional to the two-photon excitation probability, as the excitation of the core electron from the 1s to the 2s state involves a two-photon process. Therefore, the excitation probability, and consequently the HHG-XSHG yield, should scale with the square of the x-ray intensity. We investigated this dependence in our simulations, varying the x-ray intensity up to $1.4 \times 10^{17} \ \text{W/cm}^2$ while keeping the optical laser intensity of $0.56 \times 10^{14} \ \text{W/cm}^2$, as shown in Fig.~\ref{fig:depend}(b). The simulation results reveal a clear quadratic dependence on the x-ray intensity up to $8 \times 10^{16} \ \text{W/cm}^2$, indicating that the two-photon excitation process dominates in this regime. Beyond this intensity, the yields deviate from the quadratic trend, suggesting the onset of higher-order x-ray nonlinear processes.

To explore the resonance behavior of the HHG-XSHG, we performed simulations by scanning the x-ray photon energy in the range of 410 to 460 eV, while keeping the optical laser intensity of $0.56 \times 10^{14} \ \text{W/cm}^2$ and the x-ray intensity of $3.5\times10^{16} \ \text{W/cm}^2$. The results, presented in Fig.~\ref{fig:depend}(c), {\cmmtb show a peak signal at 432.3 eV which is 0.4 eV below the two-photon resonance (432.7 eV). This resonance energy shift can be explained by the laser-dressing effect to the bound state \cite{Darvasi2014,Varma2008}. When the photon energy is detuned from the laser-dressed resonance, the two-photon excitation probability and the corresponding HHG-XSHG yield drop by several orders of magnitude.} The shoulder structure on the right side of the plot is associated with the two-photon ionization process induced by the x-ray.

\subsection{Further discussion}

Realizing HHG-XSHG experimentally requires intense x-ray pulses, which can be generated by advanced XFEL lasers {\cmmtb with their short pulse durations, intense pulse energies and photon energy tunability \cite{Prat2020,Habib2023,Mcneil2010,Prat2023,Sobolev2020,Mcneil2010,Shwartz2014,Lam2018}}. In these experiments, macroscopic contributions can be crucial for achieving detectable signals, with phase matching being a key factor for efficient harmonic generation. Minimizing the phase mismatch term is essential for coherent signal accumulation, and techniques developed for laser-driven HHG can be adapted to optimize phase matching in HHG-XSHG \cite{Popmintchev2009}.

In a HHG-XSHG experiment, several competing processes can impact the signal. One significant process is ionization from valence shells, where intense optical laser pulses ionize valence electrons, shifting the two-photon resonance and introducing additional radiation pathways. This ionization can be mitigated by using lower-intensity optical pulses while still maintaining interaction with the x-ray-excited state. Moreover, Auger-Meitner decay can deplete the excited-state population, competing with HHG-XSHG. This decay occurs within a few femtoseconds (e.g., 2 to 3 fs for neon) \cite{Haynes2021}, while the rescattering process can take less than one optical cycle (approximately 2.67 fs for 800 nm) or even shorter for short trajectories (less than 1.3 fs) \cite{Corkum1993}. Consequently, Auger-Meitner decay filters out contributions from long-trajectory rescattering. Additionally, fluorescence from electrons returning to lower energy states introduces background noise near 846 eV (corresponding to the 3p to 1s transition), but this is spectrally distinct from the SHG signal.

It is crucial to highlight that the proposed laser-assisted process fundamentally differs from previously demonstrated four-wave mixing of x-rays, even in the presence of optical lasers \cite{Mukamel2005, Rouxel2020, Morillo-Candas2024, Peters2023, Rouxel2021, Alexander2024, Glover2012}. In our HHG-XSHG process, the additional optical laser plays a pivotal role by inducing strong-field excitation or facilitating tunneling ionization of x-ray-excited electrons. This leads to a highly nonlinear response to the optical field, which is distinctly different from the mechanisms driving traditional four-wave mixing. Additionally, HHG-XSHG distinguishes itself from previous XUV or x-ray-assisted HHG approaches \cite{Tross2022, Sarantseva2018, Gaarde2005, Buth2011,Tudorovskaya2014,Buth2013,Buth2013_2}. In XUV-assisted HHG, ionization typically occurs from the valence shell, with XUV pulses primarily serving to enhance conversion efficiency and extend the cutoff energy within the XUV spectral range through single-photon excitation or ionization. In contrast, our method leverages two-photon excitation and ionization of core electrons, enabling background-free detection, enhancing efficiency, and extending the emitted radiation into the x-ray spectral domain. This advancement provides a robust platform for probing ultrafast dynamics of core electrons in atoms and molecules.

The pump-probe aspect of HHG-XSHG provides powerful time-resolved capabilities for studying ultrafast electron dynamics in atomic and molecular systems. The optical laser can alter the electronic structure or drive resonant excitations, while the x-ray pulse probes these changes through the second harmonic signal, or vice versa. By adjusting the time delay between the pump and probe pulses, researchers can capture the evolution of electronic states and core-hole dynamics on their natural timescales. This capability enables the study of transient states, coherence effects, and energy transfer processes, making HHG-XSHG a versatile tool for exploring core-level dynamics and nonlinear optics in the x-ray domain.

Furthermore, broadband HHG-XSHG can generate ultrashort pulses with durations on the order of a few attoseconds, thanks to its broad spectral bandwidth, which is determined by the intensity and wavelength of the driving laser. By employing a mid-infrared driving laser, the bandwidth of the HHG-XSHG can be extended to over 1 keV, corresponding to a Fourier transform-limited pulse duration of approximately 1.8 attoseconds \cite{Popmintchev2012}. These ultrashort pulses are essential for probing valence and core electron dynamics on their intrinsic timescales, thereby opening new avenues for time-resolved studies of quantum processes and ultrafast phenomena in a variety of systems.

The photon energy range of HHG-XSHG is primarily determined by the two-photon resonance energy, which can be extended into the hard x-ray regime with photon energies exceeding 10 keV. For instance, using krypton as the interaction medium (with the 1s to 5s transition around 14 keV) and a 7 keV XFEL as the driving source.

{\cmmtb Back to the model used in the simulations, SAE approximation is widely used for solving the TDSE in strong-field and ultrafast x-ray interactions due to its computational efficiency and ability to capture essential features of HHG and related nonlinear processes \cite{Liu2020,Chen2012}. However, this approximation inherently neglects multi-electron effects, including electron-electron correlations, dynamic polarization, and collective interactions, which can significantly influence the response of atomic and molecular systems. While the SAE model remains valid for describing strong-field dynamics in atoms with a well-separated core and valence structure, its applicability becomes questionable in systems where electron correlation plays a crucial role, such as in open-shell atoms, molecules, and many-electron systems with significant configuration mixing.
Beyond the SAE approximation, multi-electron effects can be accounted for through methods such as time-dependent configuration interaction (TDCI) and time-dependent density functional theory (TDDFT) \cite{Bonafe2024,Guan2022,Liu2020}. These approaches explicitly incorporate electron correlation, which is essential for accurately describing Auger decay, shake-up and shake-off processes, and multi-electron excitations induced by intense x-ray fields. Future studies will explore these advanced techniques to improve the accuracy of our model, particularly in systems where strong correlation effects may alter the HHG-XSHG response. Additionally, including electron correlation will provide deeper insights into the role of collective excitations and correlation-driven resonances in nonlinear x-ray interactions, further advancing our understanding of attosecond x-ray spectroscopy in complex systems \cite{Jin2025}.}

\section{Conclusion and outlook}

In conclusion, we have theoretically demonstrated the feasibility of HHG-XSHG through detailed simulations on gas-phase atoms. Our results reveal a significant enhancement of the HHG-XSHG signal via two-photon excitation resonance, underscoring the crucial importance of tuning the x-ray photon energy to match core-level transitions.
This approach enables efficient HHG-XSHG generation in gases.
The HHG-XSHG mechanism combines the {\cmmtb strong-field-driven} HHG with x-ray SHG, facilitating the application of attosecond techniques developed in HHG to the x-ray regime, alongside XFEL-driven nonlinear x-ray techniques. This opens new possibilities for generating element-specific attosecond x-ray pulses and probing ultrafast core-electron dynamics and nonlinear interactions with high temporal and spectral resolution. HHG-XSHG offers a powerful tool for studying electronic structures, excited-state dynamics, and symmetry properties, positioning it as a transformative technique in attosecond science and nonlinear x-ray science.

\bibliography{xshg3}

\end{document}